\documentclass {tlp}

\usepackage {amsfonts, amssymb}

\newtheorem {theorem} {Theorem}

\newtheorem {proposition} {Proposition}
\newtheorem {corollary} {Corollary}

\newcommand {\liff} {\leftrightarrow}

\title [Decidability of the Clark's Completion Semantics for Monadic Programs
and Queries]
{Decidability of the Clark's Completion Semantics for Monadic Programs and
Queries}
\author [Levon Haykazyan] {Levon Haykazyan \\
	Mathematical Institute, University of Oxford, Woodstock Road, Oxford,
	OX2 6GG, UK \\
\email {haykazyanl@maths.ox.ac.uk}}

\submitted{3 October 2011}
\revised{15 October 2014}
\accepted{19 October 2014} 

\begin {document}

\maketitle

\begin {abstract}
There are many different semantics for general logic programs (i.e. programs
that use negation in the bodies of clauses). Most of these semantics are Turing
complete (in a sense that can be made precise), implying that they are
undecidable. To obtain decidability one needs to put additional restrictions on
programs and queries. In logic programming it is natural to put restrictions on
the underlying first-order language. In this note we show the decidability of
the Clark's completion semantics for monadic general programs and queries.
\end {abstract}

\begin {keywords}
monadic programs, general programs, Clark's completion, decidability
\end {keywords}

\section {Introduction}

{\em Definite} monadic programs have been studied by \cite {matos} and \cite
{tatsuru}. Both of these studies independently conclude that the least Herbrand
model of a monadic program is a regular set. \cite {matos} further notes that as
a consequence it is decidable whether a query follows from a monadic program.
However, if one is only interested in the decidability, then it simply follows
from the fact that monadic first order logic without equality is decidable, see
\cite {gurevich1}.

If we move from definite programs to general programs with the Clark's
completion semantics, the decidability in monadic languages does not come that
cheap - monadic logic with equality is undecidable. More precisely the
satisfiability of formulas using equality and a single monadic functional symbol
is decidable, but it becomes undecidable if formulas are allowed to use two
monadic functional symbols, see \cite {gurevich}. The central result of this
note is that the satisfiability in monadic languages becomes decidable if we
consider only models that satisfy the Clark's equational theory. As a
consequence we obtain a decidable interpreter for monadic general programs and
queries. Our proof is based on the decidability of Rabin's monadic second order
logic of successor functions \cite {rabin}.

The rest of the paper is organised as follows. In the next section we briefly go
over the preliminaries. Section 3 is devoted to detailed analysis of models of 
monadic Clark's equational theory. Next we state and prove the main result. The
final section contains some concluding remarks.

The results were obtained when the author was at Yerevan State University.

\section {Preliminaries}

In this section we recall basic definitions and results concerning logic
programs, the Clark's completion and Rabin's monadic second order logic of
successor functions.

Consider a first-order language $\cal L$. Variables are usually denoted by $x,
y, z$, constant symbols by $a, b, c$, functional symbols by $f, g, h$, predicate
symbols by $p, q, r$, terms by $s, t$, atomic formulas (or atoms) by $A, B, C$
and formulas by $F, G$ (all possibly subscripted or superscripted). A {\em
literal} is an atom (positive literal) or a negation of an atom (negative
literal). A {\em program clause} is a formula of the form
$$\forall (L_1 \land ... \land L_m \to A),$$
where $m \ge 0$, $L_1, ..., L_m$ are literals and $A$ is an atom. We abbreviate
the above clause to
$$A \gets L_1,...,L_m.$$
The atom $A$ is called the {\em head} and $L_1,...,L_m$ the {\em body} of the
clause. A {\em query} is a formula of the form
$$\exists (S_1 \land ... \land S_k),$$
where $k > 0$ and $S_1,...,S_k$ are literals. A {\em (general) program} is
a finite set of program clauses. A logic program or a query is called 
{\em definite} if all its literals are positive. A substitution $\theta =
\{x_1/t_1, ..., x_n/t_n\}$ is a finite set of pairs, where $x_i$ is a variable
and $t_i$ is a term. If $F$ is a formula, then $F\theta$ denotes the formula
obtained from $F$ by substituting all free occurrences of $x_1,...,x_n$ by
$t_1,...,t_n$ respectively.

{\em Structures} (or {\em interpretations}) are usually denoted by $\mathcal A,
\mathcal B, \mathcal M, \mathcal N$. A structure $\mathcal M$ consists of a
nonempty set $M$ and interpretations of symbols in $\mathcal L$. That is an
element $c^{\mathcal M} \in M$ for each constant symbol $c$ of $\mathcal L$, an
$n$-ary function $f^{\mathcal M} : M^n \to M$ for each $n$-ary functional symbol
$f$ of $\mathcal L$ and an $n$-ary relation $p^{\mathcal M} \subseteq M^n$ for
each $n$-ary predicate symbol of $\mathcal L$. We extend this notation to terms
of $\mathcal L$. That is if $t(x_1, ..., x_n)$ is a term, then $t^{\mathcal M}$
denotes the $n$-ary function that is the interpretation of $t$ in $\mathcal M$.

In this paper we study the most widely accepted semantics of general programs -
the Clark's completion semantics from \cite {clark}. To a program $P$ we
associate another set of formulas $c_{\cal L}(P)$ as follows. First we rewrite
each clause
$$p(t_1,...,t_n) \gets L_1,...,L_m$$
in the general form
$$p(x_1,...,x_n) \gets \exists y_1,...,y_k (x_1 = t_1 \land ... \land
x_n = t_n \land L_1 \land ... \land L_m),$$
where $x_1,...,x_n$ are new variables and $y_1,...,y_k$ are the variables of
the original clause. If
$$\begin {array} {c}
p(x_1,...,x_n) \gets E_1 \\
\vdots \\
p(x_1,...,x_n) \gets E_k
\end {array}$$
are all the general forms of clauses with $p$ in the head, then the
{\em definition} of $p$ is the formula
$$\forall (p(x_1,...,x_n) \liff E_1 \lor ... \lor E_k).$$
The empty disjunction (i.e. if $k = 0$) is understood as a logical falsehood.
It is assumed that $=$ is a new binary predicate symbol. The set of definitions
of all predicate symbols of the language is then denoted by $c_{\cal L}(P)$. The
{\em completion} of $P$, denoted by $comp_{\cal L}(P)$, is the union of $c_{\cal
L}(P)$ together with the following equality and freeness axioms referred as
$CET_{\cal L}$ (Clark's equational theory).

Equality axioms:
\begin {itemize}
\item $\forall (x = x)$;
\item $\forall (x = y \to y = x)$;
\item $\forall (x = y \land y = z \to x = z)$;
\item $\forall ((x_1 = y_1 \land ... \land x_n = y_n) \to (p(x_1,...,x_n)
\liff p(y_1,...,y_n)))$, for each predicate symbol $p$ of $\cal L$;
\item $\forall ((x_1 = y_1 \land ... \land x_n = y_n) \to f(x_1,...,x_n) =
f(y_1,...,y_n))$, for each functional symbol $f$ of $\cal L$.
\end {itemize}

Freeness axioms:
\begin {itemize}
\item $\forall (f(x_1,...,x_n) \neq g(y_1,...,y_m))$, for each pair of distinct
functional symbols $f$ and $g$ of $\cal L$ (here constants are treated as
nullary functional symbols);
\item $\forall (f(x_1,...,x_n) = f(y_1,...y_n) \to (x_1 = y_1 \land ... \land
x_n = y_n))$, for each functional symbol $f$ of $\cal L$;
\item $\forall (t(x) \neq x)$, for each term $t(x)$ of $\cal L$, where $x$ is a
proper subterm.
\end {itemize}

As the notation indicates  $comp_{\cal L}(P)$ depends not only on $P$ but also
on the underlying first order language $\cal L$. This dependence is discussed in
details in \cite {shepherdson}.  We agree to drop the language subscript,
whenever it is clear to which language we refer.

According to \cite {clark}, a logic programming system should derive
consequences of $comp(P)$ rather than $P$ itself. So given a general program $P$
and a query $Q$, the interpreter should be able to answer the following
questions:
\begin {itemize}
\item whether $comp(P) \models Q$;
\item whether $comp(P) \models \lnot Q$.
\end {itemize}
Remarkably this semantics is compatible with the widely accepted semantics of
definite logic programs. That is, for a definite program $P$ and a definite
query $Q$, we have $comp(P) \models Q$ if and only if $P \models Q$ (see \cite
{lloyd} for the proof). This, however, implies that a decidable interpreter for
general programs does not exist, since it could be used to decide whether a
definite query is a consequence of a definite program. Curiously, however, if
the language has no predicate symbols, then $CET$ itself is decidable, see e.g.
\cite {kunen}.

Next we introduce the monadic second order logic of successor functions ($SnS$)
adopted from \cite {rabin}. Briefly, $SnS$ is the monadic second order theory of
$\{1,...,n\}^*$ (finite words on $\{1, ...,n\}$) with $n$ functional symbols for
functions $x \mapsto xi$ ($i = 1,...,n$). A more precise definition follows.

The alphabet of $SnS$ consists of a countable set of object variables (usually
denoted by $x, y, z$, possibly subscripted or superscripted), a countable set
of monadic predicate variables (usually denoted by $X, Y, Z$, possibly
subscripted or superscripted), a single constant symbol $\Lambda$ and $n$ unary
functional symbols $r_1,...,r_n$, usual logical connectives, quantifiers and
punctuation symbols. Terms of $SnS$ are the usual first order terms constructed
from object variables, $\Lambda$ and $r_1,...,r_n$. Formulas of $SnS$ are
defined as follows:
\begin {itemize}
\item if $t, s$ are terms and $X$ is a predicate variable, then $t = s$ and
$X(t)$ are (atomic) formulas ($X(t)$ is also written as $t \in X$);
\item if $F, G$ are formulas, $x$ is an object variable and $X$ is a predicate
variable, then $(\lnot F)$, $(F \land G)$, $(F \lor G)$, $(F \to G)$, $(F \liff
G)$, $(\exists x F)$, $(\exists X F)$, $(\forall x F)$, $(\forall X F)$ are
formulas.
\end {itemize}

The semantics of $SnS$ formulas is defined with respect to the term
interpretation and the usual second order semantics. That is consider the
structure ${\cal U}_{SnS}$ whose domain is the set of ground terms.  The
constant symbol $\Lambda$ is interpreted by the ground term $\Lambda$ and each
functional symbol symbol $r$ is interpreted by the function $t \mapsto r(t)$.
Given an $SnS$ sentence $F$, the relation ${\cal U}_{SnS} \models F$ is defined
as in the standard second order semantics. That is object quantifiers range over
the domain, and predicate quantifiers range over all subsets of the domain. We
usually surpass ${\cal U}_{SnS}$ from the notation and say that a sentence $F$
is true if ${\cal U}_{SnS} \models F$. The decidability of $SnS$ is crucial for
our purposes.

\begin {theorem} [see \cite {rabin}]
There is an algorithm for deciding if a given $SnS$ formula is true.
\end {theorem}

From now on we fix a finite monadic language $\cal L$. 

\section {Models of Monadic Clark's Equational Theory}

It is well known that in the study of theories that contain equality axioms, one
can restrict attention to structures where $=$ is interpreted as the equality in
the domain. So without loss of generality, we will assume that in all structures
$=$ is interpreted as the equality. This further ensures that equality axioms of
$CET$ hold. So $CET$ is reduced to freeness axioms only.

Let $\cal M$ be a structure with domain $M$.  An element $a \in M$ is said to
{\em proceed} $b \in M$ if there is a term $t(x)$ containing the variable $x$
such that $t^{\cal M}(a) = b$. A {\em root} element is an element that does not
have predecessors, apart from itself. Let ${\cal L}_0$ denote the language
consisting of only the functional symbols of $\cal L$.

\begin {proposition}
A structure is a model of $CET_{\cal L}$ if and only if it is a model of
$CET_{{\cal L}_0}$ and each constant symbol is interpreted as a distinct root
element.
\end {proposition}

\begin {proof}
Let $\cal A$ be a model of $CET_{\cal L}$. Then $\cal A$ is a model of
$CET_{{\cal L}_0}$ since $CET_{{\cal L}_0} \subseteq CET_{\cal L}$. Further,
$$\forall x (f(x) \neq a)$$
is in $CET_{\cal L}$ for each functional symbol $f$ and each constant symbol
$a$. So each constant symbol is interpreted as a root element. Since
$$a \neq b$$
is in $CET_{\cal L}$ for distinct constant symbols $a$ and $b$, each constant
symbol is interpreted as a distinct element.

Conversely let $\mathcal A$ be a model of $CET_{\mathcal L_0}$ where each
constant symbol is interpreted by a distinct root element. Note that $CET_{\cal
L}$ is obtained from $CET_{{\cal L}_0}$ by adding axioms
$$\forall x (f(x) \neq a)$$
for each functional symbol $f$ and each constant symbol $a$ and
$$a \neq b$$
for each distinct constant symbols $a$ and $b$.
All these axioms hold in $\cal A$, so it is a model of $CET_{\cal L}$.
\end {proof}

With this characterisation in mind, let us study structures in ${\cal L}_0$. Let
$\{{\cal M}_i : i \in I\}$ be a set of structures in ${\cal L}_0$. We can
define their disjoint union $\coprod_{i \in I} {\cal M}_i$ as the structure
whose domain is the disjoint union of domains of ${\cal M}_i$ and each
functional symbol $f$ is interpreted as $f^{{\cal M}_i}$ in the domain of ${\cal
M}_i$.

\begin {proposition}
The structure $\coprod_{i \in I} {\cal M}_i$ is a model of $CET_{{\cal L}_0}$
if and only if each ${\cal M}_i$ is.
\end {proposition}

\begin {proof}
The structure ${\cal M}_i$ is a substructure of $\coprod_{i \in I} {\cal M}_i$.
So, if the latter is a model of $CET_{{\cal L}_0}$, then so is the former since
$CET_{{\cal L}_0}$ is a universal theory.

Conversely assume that ${\cal M}_i \models CET_{{\cal L}_0}$ for each $i \in I$.
Let $f$ and $g$ be distinct functional symbols. If $x$ and $y$ belong to
the domains of different structures than $f(x) \neq g(y)$ holds in $\coprod_{i
\in I} {\cal M}_i$. If $x$ and $y$ belong to the domain of ${\cal M}_i$, then
$f(x) \neq g(y)$ holds in ${\cal M}_i$ and hence in $\coprod_{i \in I} {\cal
M}_i$. So
$$\coprod_{i \in I} {\cal M}_i \models \forall (f(x) \neq g(y)).$$
Other axioms of $CET_{{\cal L}_0}$ are checked similarly.
\end {proof}

Now let $\cal M$ be a model of $CET_{{\cal L}_0}$. Two elements $a, b \in M$ are
called {\em connected} (in symbols $a \sim b$) if they have a common
predecessor.

\begin {proposition}
The relation $\sim$ is an equivalence relation.
\end {proposition}

\begin {proof}
It is easy to see that $\sim$ is reflexive and symmetric. For transitivity let
$a_1 \sim a_2$ and $a_2 \sim a_3$. Let $b$ be the common predecessor of $a_1$
and $a_2$ and $c$ be the common predecessor of $a_2$ and $a_3$. Then there
exist terms $t(x) = h_1(...h_k(x)...)$ and $s(x) = g_1(...g_l(x)...)$ such that
$t^{\cal M}(b) = a_2 = s^{\cal M}(c)$. Without loss of generality assume that 
$l \le k$. Then by freeness axioms $h_1 = g_1, ..., h_l = g_l$ and $c =
h_{l+1}^{\cal M}(...h_k^{\cal M}(b)...)$. Thus $b$ proceeds $c$ and hence $a_3$
and so $a_1$ and $a_3$ are connected.
\end {proof}

Thus $\sim$ partitions $M$ into equivalence classes.  Each class is closed under
the interpretations of the functional symbols and so generates a substructure.
We will refer to these substructures as the {\em components} of $\cal M$. Thus
$\cal M$ is isomorphic to the disjoint union of all of its components. Our goal
is to characterise each component.

There can be at most one root element in each component. We will refer to
components containing root elements as {\em root components}. If $\cal A$ is a
root component with domain $A$ and root element $a$, then $A = \{t^{\cal A}(a) :
t(x) \hbox { is a term containing } x\}$. Further, by freeness axioms, elements
$t^{\cal A}(a)$ are all different for different terms $t(x)$. Thus the
substructure generated by $A$ is isomorphic to the term structure of $\langle c,
f_1, ..., f_n \rangle$, where $c$ is some constant symbol and $f_1, ..., f_n$
are the functional symbols of ${\cal L}_0$. We will call this structure the {\em
root structure}. Thus all root components are isomorphic to the root structure
(and hence are isomorphic to each other).

Now let us study components that do not contain a root element - {\em non-root
components}. Let $\cal A$ be a non-root component with domain $A$. Pick
arbitrary $a_0 \in A$. Then there are $a_1 \in A$ and a functional symbol $h_1$
such that $a_0 = h_1^{\cal A}(a_1)$. Similarly there are $a_2 \in A$ and a
functional symbol $h_2$ such that $a_1 = h_2^{\cal A}(a_2)$. Continuing this way
we will get an infinite sequence $a_0, a_1, a_2, ...$ of elements of $A$ and an
infinite sequence of functional symbols $h_1, h_2, ...$. By freeness axioms all
$a_j$ are different. Observe that $\{a_0, a_1, ... \}$ is the set of
predecessors of $a_0$.  For an arbitrary $b \in A$, elements $b$ and $a_0$
should have a common predecessor. Let $a_j$ be the one with the minimal index
and let $b = g_1^{\cal A}(...g_k^{\cal A}(a_j)...)$. In case that $j > 0$ and $k
> 0$ we would further have $g_k \neq h_j$. Let $c_0, c_1, ...$ be new constant
symbols, $f_1, ..., f_n$ be the functional symbols of ${\cal L}_0$ and consider
the set of ground terms over $\langle c_0, c_1, ..., f_1, ..., f_n \rangle$ that
do not contain $h_j(c_j)$ as a subterm for $j = 1, 2, ...$. Define the
interpretations of functional symbols as
$$f(t) = \left\{
\begin {array} {l}
c_{j - 1} \mbox { if $t = c_j$ and $f = h_j$ for some $j = 1, 2, ...$} \\
f(t) \mbox { otherwise}
\end {array}
\right.$$
From the above discussion it follows that this structure is isomorphic to $\cal
A$. We will refer to such structures as {\em non-root structures}. The sequence
$h_1, h_2, ...$ is called the signature
of the structure. Note that non-root structures with different signatures
may be isomorphic. To sum up, we obtain the following

\begin {theorem}
A structure is a model of $CET$ if and only if it is a disjoint union of root
structures and non-root structures and each constant symbol is interpreted as a
root element.
\end {theorem}

\section {Decidability for Monadic Programs and Queries}

In this section we construct an algorithm to decide whether a monadic query (or
its negation) is a consequence of $comp(P)$ for a monadic program $P$. In
fact we show slightly more: given an arbitrary monadic formula $F$ (probably
using equality) it is decidable whether $\{F\} \cup CET$ is consistent or not.
Let $c_1, ..., c_k$ be the constant, $f_1, ..., f_n$ - the functional and $p_1,
..., p_m$ - the predicate symbols of $\cal L$. We will construct a formula of
$S(2n + 1)S$ that would be true if and only if $\{F\} \cup CET$ is consistent.
For convenience we will refer to the functional symbols of $S(2n + 1)S$ as $f_0,
f_1, ..., f_n, f_1^{-1}, ..., f_n^{-1}$. This is a bit confusing since for
positive $j$ we also use $f_i^j(t)$ to denote the term $f_i(...f_i(t)...)$,
where $f_i$ is repeated $j$ times. However, the notation $f_i^{-1}$ indicates
exactly how we are going to use that functional symbol.

Let $D$ be a subset of $U_{S(2n+1)S}$ such that 
\begin {itemize}
\item $f_0^{j}(\Lambda) \in D$ for $j = 1, ..., k$;
\item for every $x \in D$ and every $i = 1, ..., n$ either $f_i(x) \in D$ or $x
= f_i^{-1}(y)$ for some $y \in D$, but not both. 
\end {itemize}
Let $P_1, ..., P_m$ be subsets of $D$. The tuple $\langle D, P_1, ..., P_m
\rangle$ defines an interpretation of $\cal L$ in the following way:
\begin {itemize}
\item the domain of the interpretation is $D$;
\item the constant symbol $c_j$ is interpreted as $f_0^{j}(\Lambda)$;
\item the functional symbol $f_i$ is interpreted as the function
$$x \mapsto \left\{
\begin {array} {l}
f_i(x) \mbox { if } f_i(x) \in D \\
y \mbox { if } x = f_i^{-1}(y);
\end {array}
\right.$$
\item the predicate symbol $p_l$ is interpreted as the set $P_l$.
\end {itemize}
We want to find and express in $S(2n+1)S$ sufficient conditions on $D$, such
that structures defined by $\langle D, P_1, ..., P_m \rangle$ enumerate
countable models of $CET$ and only those. 

Let $domain(X)$ denote the following $S(2n+1)S$ formula:
$$\begin {array} {ll}
& \bigwedge_{j = 1, ..., k} f_0^{j}(\Lambda) \in X \\
\land & \\
& \forall x (x \in X \to \bigwedge_{i = 1, ..., n} (f_i(x) \in X \veebar
\exists y \in X~x = f_i^{-1}(y))) \\
\land & \\
& \bigwedge_{j = 1, ..., k \atop i = 1, ..., n} f_i^{-1}(f_0^{j}(\Lambda)) \not
\in X
\\
\land & \\
& \bigwedge_{i = 1, ..., n} \forall x (x \in X \land f_i(x) \in X \to
\bigwedge_{i' = 1, ..., n} f_{i'}^{-1}(f_i(x)) \not \in X) \\
\land & \\
& \forall x (\lnot \bigvee_{i = 1, ..., n-1 \atop i' = i+1, ..., n} (f_i^{-1}(x)
\in X \land f_{i'}^{-1}(x) \in X)),
\end {array} $$
where $\veebar$ stands for the exclusive or.

\begin {proposition}
\label {from}
If $domain(D)$ holds for $D \subseteq U_{S(2n+1)S}$ and $P_1, ..., P_m
\subseteq D$ then the interpretation defined by $\langle D, P_1, ..., P_m
\rangle$ is a model of $CET$.
\end {proposition}

\begin {proof}
Let $D$ be a subset of $U_{S(2n+1)S}$ such that $domain(D)$ holds and $P_1, ...,
P_m \subseteq D$. The first two clauses of the definition of $domain$ ensure
that $\langle D, P_1, ..., P_m \rangle$ defines a structure $\cal D$. Let us
show that it satisfies $CET$. By the third clause $f_0^{j}(\Lambda)$ is a root
element for $j = 1, ..., k$. So we need to only check the axioms of $CET_{{\cal
L}_0}$.

Let $a, b \in D$ and assume that $f_i^{\cal D}(a) = f_{i'}^{\cal D}(b)$.
Consider two cases. 
\begin{itemize}
\item If $f_i(a) \in D$, then $f_{i'}^{\cal D}(b) = f_i^{\cal D}(a) = f_i(a)$.
But then by the fourth clause $f_{i'}^{-1}(f_i(a)) \not \in D$ and so $b \neq
f_{i'}^{-1}(f_i(a))$. It follows that $i = i'$ and $a = b$.
\item If $f_i(a) \not \in D$, then $a = f_i^{-1}(c)$ for some $c \in D$. Thus we
have $f_{i'}^{\cal D}(b) = c$. Then $f_{i'}(b) \not \in D$, since otherwise $a =
f_i^{-1}(f_{i'}(b)) \in D$ contrary to the fourth clause. But then
$f_{i'}^{-1}(c) = b \in D$. By the fifth clause $i = i'$ and $a = b$.
\end{itemize}
Thus in both cases we have $i = i'$ and $a = b$. Therefore
$${\cal D} \models \forall(f_i(x) \neq f_{i'}(y)),$$
for $i \neq i'$ and
$${\cal D} \models \forall(f_i(x) = f_i(y) \to x = y).$$

To show that the third axiom scheme of $CET_{{\cal L}_0}$ holds assume that $a
\in D$ and $f_{n_1}^{\cal D}(...f_{n_i}^{\cal D}(a)...) = a$. Denote $b =
f_{n_2}^{\cal D}(...f_{n_i}^{\cal D}(a)...)$, so that $f_{n_1}^{\cal D}(b) =
a$. Again consider two cases
\begin {itemize}
\item Assume $a = f_{n_1}(b)$. But then for no $a' \in D$, $a = 
f_{n_i}^{-1}(a')$. So $f_{n_i}(a) \in D$. Similarly $f_{n_{i-1}}(f_{n_i}(a)) \in
D$ and continuing this way we will get that $f_{n_1}(...f_{n_i}(a)...) \in D$.
So $f_{n_1}^{\cal D}(...f_{n_i}^{\cal D}(a)...) = f_{n_1}(...f_{n_i}(a)...) \neq
a$, which contradicts our assumption.
\item Otherwise $b = f_{n_1}^{-1}(a)$. But we have $b = f_{n_2}^{\cal
D}(...f_{n_i}^{\cal D}(a)...)$. So $f_{n_3}^{\cal D}(...f_{n_i}^{\cal D}(a)...)
= f_{n_2}^{-1}(b) = f_{n_2}^{-1}(f_{n_1}^{-1}(a))$. Continuing this way we will
get $a = f_{n_i}^{-1}(...f_{n_1}^{-1}(a)...)$ which is not possible.
\end {itemize}
Thus in both cases we obtain a contradiction, which proves that the third axiom
scheme of $CET_{{\cal L}_0}$ holds.
\end {proof}

The last proposition ensures that whenever $domain(D)$ holds, $D$ defines a
model of $CET$. We also need each countable model of $CET$ to have such a
representation.

\begin {proposition}
\label {to}
For every countable model $\cal D$ of $CET$, there are $D \subseteq
U_{S(2n+1)S}$ and $P_1, ..., P_m \subseteq D$ such that $domain(D)$ holds and
the structure defined by $\langle D, P_1, ..., P_m \rangle$ is isomorphic to
$\cal D$.
\end {proposition}

\begin {proof}
Let $\cal D$ be a countable model of $CET$. Let $\cal D$ be obtained by
interpreting $c_1, ..., c_k$ as root elements in $\coprod_{j \in J} {\cal D}_j$,
where ${\cal D}_j$ is either the root structure or some non-root structure.
Since $\cal D$ is countable, $J$ is at most countable. So without loss of
generality we can assume that $J \subseteq \mathbb N$. We can also assume that
$\{1, ..., k\} \subseteq J$ and that $c_1, ..., c_k$ are interpreted as the root
elements of ${\cal D}_1, ..., {\cal D}_k$. Since $P_1, ..., P_m$ can be chosen
arbitrarily, it is enough to find $D \subseteq U_{S(2n+1)S}$ such that
$domain(D)$ holds and the structure generated by $D$ is ${\cal L}_0$-isomorphic
to $\coprod_{j \in J} {\cal D}_j$ (since $domain(D)$ holds $f_0^j(\Lambda)$
is a root element for $j = 1, ... ,k$).

A $S(2n+1)S$ term $t(x)$ is called a {\em main} term if it contains $x$ and
does not contain $f_0$. For $a \in U_{S(2n+1)S}$, define the subtree rooted in
$a$ as the set $T(a) = \{t(a) : t(x) \hbox { is a main term} \}$. We will
represent $D$ as a union $D = \bigcup_{j \in J} D_j$, where $D_j \subseteq
T(f_0^j(\Lambda))$. If ${\cal D}_j$ is a root structure, the choice of $D_j$ is
straightforward: $D_j = \{t(f_0^j(\Lambda)) : t(x) \hbox { is a term over } f_1,
..., f_n \hbox { and } x\}$. Clearly $D_j$ generates a root structure with
$f_0^j(\Lambda)$ as the root element.

Now let ${\cal D}_j$ be a non-root structure over $\langle d_0, d_1, ..., f_1,
..., f_n \rangle$ and $f_{n_1}, f_{n_2}, ...$ be its signature. Thus the domain
of ${\cal D}_j$ consists of ground terms not containing $f_{n_i}(d_i)$ as
subterms for $i = 1, 2, ...$ and $f_{i'}$ is interpreted as $t \mapsto
f_{i'}(t)$ with the exception that $f_{n_i}(d_i) = d_{i-1}$. To form $D_j$ we
pick the element
$f_{m_1}(...f_{m_l}(f_{n_i}^{-1}(...f_{n_1}^{-1}(f_0^j(\Lambda))...))...)$ for
the element $f_{m_1}(...f_{m_l}(d_i)...)$ of ${\cal D}_j$. Denote the structure
generated by $D_j$ as ${\cal D}_j'$. Note that $f_{n_i}^{{\cal
D}_j'}(f_{n_i}^{-1}(...f_{n_1}^{-1}(f_0^j(\Lambda))...)) =
f_{n_{i-1}}^{-1}(...f_{n_1}^{-1}(f_0^j(\Lambda))...)$ (which corresponds to
$f_{n_j}(d_j) = d_{j-1}$) and $f_i^{{\cal D}_j'}(t) = f_i(t)$ otherwise. This
shows that ${\cal D}_j$ and ${\cal D}_j'$ are indeed isomorphic.

It is routine to check that $D$ satisfies $domain(X)$.
\end {proof}

Last two propositions enable us to quantify over all countable models of $CET$.
So to decide whether a formula $F$ has a model satisfying $CET$ we need to find
an $S(2n+1)S$ formula to define the models of $F$. The formula $F$ is called
{\em simple} if every functional symbol $f$ occurs in a subformula of the form
$y = f(x)$.

\begin {proposition}
\label {true}
For a simple closed formula $F$, there is an $S(2n+1)S$ formula $Mod_F(X,
Y_1, ..., Y_m)$ such that for every $D \subseteq U_{S(2n+1)S}$ satisfying
$domain(X)$ and every $P_1, ..., P_m \subseteq D$ the following holds: $Mod_F(D,
P_1, ..., P_m)$ holds if and only if the structure defined by $\langle D, P_1,
..., P_m \rangle$ is a model of $F$.
\end {proposition}

\begin {proof}
Without loss of generality assume that $F$ uses only the connectives $\lor$ and
$\lnot$ and the quantifier $\exists$. To obtain $Mod_F(X, Y_1, ..., Y_m)$ we do
the following
\begin {itemize}
\item replace each subformula $\exists x G$, with $\exists x (x \in X \land G)$
\item replace each predicate symbol $p_l$ with a predicate variable $Y_l$;
\item replace each constant symbol $c_j$ by the term $f_0^j(\Lambda)$;
\item replace each subformula of the form $y = f(x)$ with $y = f(x) \lor x =
f^{-1}(y)$.
\end {itemize}
Now let $D \subseteq U_{S(2n+1)S}$ be such that $domain(D)$ holds and $P_1, ...,
P_m \subseteq D$. Denote by $\cal D$ the structure defined by $\langle D, P_1,
..., P_m \rangle$. Consider an arbitrary simple $\cal L$-formula $G$ (possibly
with free variables). Let $\phi$ be an assignment of its free variables (with
respect to $\cal D$). Since the range of $\phi$ is in $U_{S(2n+1)S}$ (the set of
$S(2n+1)S$ ground terms), it defines an $S(2n+1)S$ substitution. We show by
induction on the construction of $G$ that ${\cal D} \models_\phi G$ if and only
if $Mod_G(D, P_1, ..., P_m)\phi$ holds.
\begin {itemize}
\item If $G$ is $p_l(t)$, then $t$ does not contain functional symbols. If $t$
is a variable $x$, then ${\cal D} \models_\phi p_l(x) \iff \phi(x) \in P_l$ by
definition of ${\cal D}$. Otherwise $t$ is a constant symbol $c_j$ and then
${\cal D} \models_\phi p_l(c_j) \iff f_0^j(\Lambda) \in P_l$ again by definition
of ${\cal D}$.
\item If $G$ is $t_1 = t_2$, then consider two cases. If $t_1$ and $t_2$ do not
contain functional symbols, then ${\cal D} \models_\phi t_1 = t_2 \iff Mod_G(D,
P_1, ..., P_m)\phi$ can be shown similar to the previous case. Otherwise $G$ is
of the form $y = f_i(x)$. In this case we have ${\cal D} \models_\phi y = f_i(x)
\iff \phi(y) = f_i^{\cal D}(\phi(x)) \iff \phi(y) = f_i(\phi(x)) \lor \phi(x) =
f_i^{-1}(\phi(y))$.
\item The cases $G = G_1 \lor G_2$ and $G = \lnot G_1$ are completely
straightforward.
\item If $G$ is $\exists x G_1$, then ${\cal D} \models_\phi \exists x G_1 \iff$
for some $t \in D$, ${\cal D} \models_{\phi[x \mapsto t]} G_1 \iff$ for some $t
\in D$, we have $Mod_{G_1}(D, P_1, ..., P_m)\phi[x \mapsto t] \iff \exists x (x
\in D \land Mod_{G_1}(D, P_1, ..., P_m)\phi \setminus x)) \iff Mod_G(D, P_1,
..., P_m)\phi$. Here $\phi[x \mapsto t]$ and $\phi \setminus x$ are
substitutions that differs from $\phi$ only in assignment of $x$. The
substitution $\phi[x \mapsto t]$ assigns $t$ to $x$ and $\phi \setminus x$ does
not assign anything to $x$.
\end {itemize}
Now since $F$ does not contain free variables, ${\cal D} \models F \iff
Mod_F(D, P_1, ..., P_m)$.
\end {proof}

It remains to glue all the pieces together.

\begin {theorem}
There is an algorithm that takes a finite monadic language $\cal L$ and a
formula $F$ in $\cal L$ (possibly using equality) and decides whether $\{F\}
\cup CET_{\cal L}$ is satisfiable.
\end {theorem}

\begin {proof}
First we transform $F$ into $F'$ by repeatedly replacing each atomic subformula
$A(f(t))$ not of the form $y = f(x)$ by $\exists x, y (x = t \land y =
f(x) \land A(y))$ until there is none left. Clearly $F'$ is simple and is
logically equivalent to $F$. Then we form $Mod_{F'}$. By Propositions \ref
{from} and \ref {to}, the formula $domain(X)$ enumerates all countable models of
$CET_{\mathcal L}$ and by Proposition \ref {true}, the formula $Mod_{F'}(X, Y_1,
..., Y_m)$ defines the truth of $F'$ in $\langle X, Y_1, ..., Y_m \rangle$.
Hence $\{F\} \cup CET_{\cal L}$ is satisfiable if and only if the $S(2n+1)S$
formula 
$$\exists X, Y_1, ..., Y_m (domain(X) \land Y_1
\subseteq X \land ... Y_m \subseteq Y_m \land Mod_{F'}(X, Y_1, ..., Y_1))$$ 
is true, which is decidable.
\end {proof}

\begin {corollary}
There is an algorithm that given a finite monadic language $\cal L$, a program
$P$ and a query $Q$ decides the following questions
\begin {itemize}
\item whether $comp_{\cal L}(P) \models Q$;
\item whether $comp_{\cal L}(P) \models \lnot Q$.
\end {itemize}
\end {corollary}

\section {Conclusion}

The precise computational complexity of the decision procedure for the Clark's
completion semantics remains to be determined. The decision procedure for $SnS$
is primitive recursive, but not elementary recursive, (i.e. its complexity
cannot be bound by a tower of exponentials of a fixed length) see \cite {meyer}.
This makes the proposed algorithm for deciding the Clark's completion semantics
prohibitive for practical applications. For comparison exponential algorithms
are known for deciding the satisfiability of a monadic first order formula
without equality and the satisfiability of a monadic first order formula with
equality but without functional symbols \cite {gurevich2}.

\bibliographystyle{acmtrans}
\bibliography {ref}

\end {document}